\documentclass[pra,aps,twocolumn,showpacs]{revtex4}
\usepackage{graphicx}

\begin{document}
\draft \pagenumbering{roma}
\title{Realization of symmetric sharing of entanglement
in semiconductor microcrystallites coupled by a cavity field}

\author{Yu-xi Liu,$^{1}$   Adam Miranowicz,$^{1,2,3}$
Masato Koashi,$^{1,2}$ and  Nobuyuki Imoto$^{1,2,4}$}

\date{12 December 2002; published in Physical Review A {\bf 66}, 062309 (2002)}

\address{
$^1$The Graduate University for Advanced Studies (SOKENDAI),
Shonan Village, Hayama, Kanagawa 240-0193, Japan\\
$^2$CREST Research Team for Interacting Carrier Electronics,
Hayama, Kanagawa 240-0193, Japan\\
$^3$Nonlinear Optics Division, Institute of Physics, Adam
Mickiewicz University, 61-614 Pozna\'n, Poland\\
$^4${NTT Basic Research Labs,~3-1 Morinosato-Wakamiya, Atsugi,
Kanagawa 243-0198, Japan}\\}

\begin{abstract}
The entanglement of excitonic states in a system of  $N$ spatially
separated semiconductor microcrystallites is investigated. The
interaction among the different microcrystallites is mediated by a
single-mode cavity field. It is found that the symmetric sharing
of the entanglement (measured by the concurrence) between any pair
of the excitonic state with $N$ qubits defined by the number
states (vacuum and a single-exciton states) or the coherent states
(odd and even coherent states) can be prepared by the cavity field
for this system.

\end{abstract}


\maketitle

\pagenumbering{arabic}

\section{INTRODUCTION}
Quantum entanglement plays an important role in the quantum
communication and quantum-information processing. One can
implement quantum teleportation of a given state from one place to
another by virtue of the entangled state~\cite{Zeilinger,AB}.
Entanglement among many particles is essential for most quantum
communication schemes. The simplest generalization of the
entangled states for more than two particles is the so-called GHZ
state~\cite{ghz}. The entanglement of the nearest-neighbor pairs
for  an infinite collection of qubits arranged in a line was
studied by Wootters~\cite{wootters}. For a finite system, Koashi
{\em et al.}~\cite{koashi}  investigated the completely symmetric
sharing of entanglement for an arbitrary pair of $N$ qubits. D\"ur
studied not only the symmetric sharing  of the pairwise
entanglement but also the nonsymmetric sharing in a system of $N$
qubits~\cite{dur}. The nearest-neighbor entanglement of $N$ qubits
in a ring configuration was also studied and further a concrete
physical system of $N$ spin-$\frac{1}{2}$ particles interacting
via the Heisenberg antiferromagnetic Hamiltonian was
given~\cite{kwk}. The question arises whether it is possible or
not to prepare states with the symmetric sharing of entanglement
in some real systems within the present-day technology. And, if
yes, how to achieve such entangled states?

Any many-body system with defined qubits, if set in a properly
chosen state, will evolve through states containing entangled
qubits. Similarly, most of the ground states of real systems
contain entangled states. However, for the purposes of quantum
computation and quantum information, the most important aspects of
quantum entanglement are especially
(i) deterministic control over the quantum coherence of states and
 (ii) time evolution and occurrence of maximally entangled states.
In this study, we focus on the latter topic, specifically, on the
generation of the maximal pairwise entanglement.  As was shown by
Koashi et al.~\cite{koashi} that entanglement cannot be
unlimitedly shared among an arbitrary number of qubits and the
degree of bipartite entanglement decreases with the increasing
number of entangled pairs in an $N$-qubit system in which any pair
of particles is entangled. The maximum degree of bipartite
entanglement, measured in the concurrence, between any pair of
qubits is bound by $2/N$. Here, we will investigate a physical
realization of this maximally possible bipartite entanglement.

Within the past few years,  advances in microfabrication
technology have allowed researchers to create unique quantum
confinement, and thereby have opened up a new realm of fundamental
physics. As low-dimensional semiconductor structures, quantum dots
attract a considerable interest because of their atomlike
properties. They can lead to novel optoelectronic devices that can
be applied to the emerging fields of quantum
computing~\cite{computer,Ima} and quantum-information
processing~\cite{information,molo}. It is well known that
Coulomb-correlated electron-hole pairs called excitons can be
optically  generated and controlled in a single dot~\cite{ed}, and
thus can be used to store the quantum information and realize
quantum computing~\cite{Joh}. On the other hand, a significant
fraction of quantum computing and information schemes relies on
the strong-coupling regime of the cavity quantum electrodynamics
(QED). The observed Rabi oscillations of excitons in a single
quantum dot~\cite{kam} suggest the possibility that the quantum
dot cavity QED will be realized in the near future. However, an
essential feature of a quantum dot is that the electronic energy
levels are completely quantized, so the behavior of excitons
deviates from the bosonic statistics. In the present paper, we
will consider some slightly bigger semiconductor microstructures,
such as the microcrystallites. In this case, the area of the
microcrystallite is larger than that of the Bohr radius of the
exciton, and the behavior of the excitons with low excitation are
the same as that of the bosonic particles. Chuang {\em et
al.}~\cite{Yama} showed that the quantum code of the bosonic mode
enables a more efficient error correction. So, the mode of the
excitons offers a possible physical implementation for such
bosonic-mode coding. We propose a possible scheme to prepare the
entangled excitonic states for the symmetric sharing in the system
of $N$ microcrystallites by virtue of the cavity QED. The cavity
field mediates the interaction among semiconductor
microcrystallites, and then the entangled excitonic states can be
prepared by the cavity field.

We organize our paper as follows. In Sec. II, we will propose a
scheme based on the present-day technology and model the
Hamiltonian of the whole system. The solutions corresponding to
this Hamiltonian are given for the general initial state. In Secs.
III and IV, the bosonic exciton operator is used as an approach to
deal with qubits uniformly. We will show how to prepare the
entangled excitonic states with  qubits defined by  different
excitonic states using various initial conditions of the cavity
field. Any physical system cannot be isolated from its
environment. The interaction between the system and  the
environment will result in their entanglement, then coherence of
the qubit is destroyed with the time evolution. So, in Sec. V, we
will demonstrate the environment effect on the entangled states.
Finally, some comments and conclusions will be given.

\section{MODEL AND ITS SOLUTION}
We assume that there are $N$ spatially separated semiconductor
microcrystallites (also called large semiconductor quantum
dots~\cite{Han,Ban,Eng}) which are placed into an ideal
semiconductor microcavity with a single-mode field, for example,
the microcrystallites are embedded in a disk structure of the
semiconductor, which is similar to the Imamo\v{g}lu model for
quantum dots~\cite{Ima}. And we assume that the radius $R$ of each
microcrystallite is much larger than the Bohr radius $a_{B}$ of
excitons, but smaller than the wavelength $\lambda$ of the cavity
field, that is, $a_{B}\ll R \le \lambda$. Also the distance
between each pair of microcrystallites is larger than the optical
wavelength $\lambda$ of the cavity field, and the
microcrystallites indirectly interact by virtue of the cavity
field. We also assume that there are few electrons excited from
the valence band to the conduction band such that the exciton
density for each microcrystallite is much smaller than the Mott
density. So, all nonlinear terms included in the interaction of
the exciton-exciton and exciton-photon can be neglected in our
model, and the excitons are considered as ideal bosons. The cavity
field is assumed to resonantly interact with the zero-momentum
excitons in each microcrystallite, the thermalization of the
excitons is neglected. Under the above conditions, we can use the
effective Hamiltonian under the rotating wave approximation as
follows~\cite{Han,Ban}
\begin{equation}
H=\hbar\omega a^{\dagger}a+\hbar\omega\sum_{j=1}^{N}
b_{j}^{\dagger}b_{j} +\hbar \sum_{j=1}^{N}g_{j}(a^{\dagger}b_{j}+a
b^{\dagger}_{j}),
\end{equation}
where $a(a^{\dagger})$ is the annihilation (creation) operator of
the cavity field with  frequency $\omega$, and
$b_{j}(b_{j}^{\dagger})$ is the annihilation (creation) operator
of the excitons in the $j$th microcrystallite with the same
frequency $\omega$ as that of the cavity field. First, we assume
that the coupling constants $g_{j}$ with $j=1, \hspace{0.1cm}
2,\hspace{0.1cm} \cdots,\hspace{0.1cm} N$ between the cavity field
and microcrystallites are different. We can give the Heisenberg
equations of motion for the operators of the cavity field  and the
excitons as follows
\begin{mathletters}
\begin{eqnarray}
\frac{\partial A(t)}{\partial t}&=&-i \sum_{j}g_{j} B_{j}(t),\label{eq:2a}\\
\frac{\partial B_{j}(t)}{\partial t}&=&-i g_{j} A(t)
\hspace{0.3cm}(j=1, 2, \cdots, N) \label{eq:2b},
\end{eqnarray}
\end{mathletters}
where the transformations $a(t)=A(t)e^{-i\omega t}$ and
$b_{j}(t)=B_{j}(t)e^{-i\omega t}$ are applied. The solutions of
Eqs.~(\ref{eq:2a}) and (\ref{eq:2b}) can be obtained  as
\begin{mathletters}
\begin{eqnarray}
A(t) &=&a(0)\cos (G^{\prime}t)-i\sum_{j}f_{j}b_{j}(0), \\
B_{j}(t)
&=&\sum_{m}\left\{\delta_{jm}+\frac{g_{j}g_{m}
[\cos(G^{\prime}t)-1]}{G^{\prime
2 }}\right\} b_{m}(0)\nonumber\\
&-&if_{j}a(0), \label{eq:3b}
\end{eqnarray}
\end{mathletters}
\noindent where $G^{\prime}\equiv \sqrt{\sum_{j=1}^{N}g^2_{j}}$;
$f_{j}\equiv f_{j}(t)=g_{j}\sin (G^{\prime}t)/G^{\prime}$, $a(0)$
and $b_{j}(0)$  $(j=1,\cdots, N)$ are the initial operators of the
cavity field and excitons, respectively. We assume that the
initial state of the whole system is $|\Psi(0)\rangle=
|\psi(0)\rangle_{C} |0\rangle^{\otimes N}$, which means that the
cavity field is initially in the state $|\psi(0)\rangle_{C}$, but
there is no exciton in any microcrystallite. Then we can obtain
the wave function as follows
\begin{equation}
|\Psi(t)\rangle=U(t)|\psi(0)\rangle_{C}|0\rangle^{\otimes N},
\label{eq:5}
\end{equation}
with the time-evolution operator $U(t)=e^{-iHt/\hbar}$.

\section{PREPARATION OF ENTANGLED EXCITONIC STATE BY SINGLE-PHOTON STATE}

\begin{figure}
\includegraphics[width=8cm]{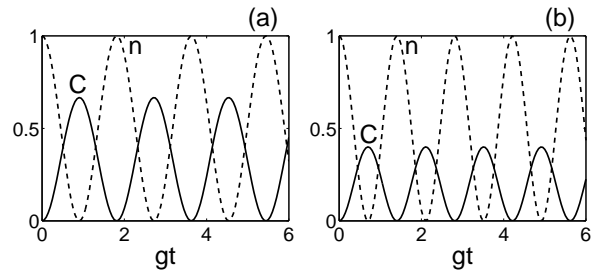}
\caption[]{Time evolutions  of both the concurrence $C$ (solid
line) and the average photon number $n=\langle
a^{\dagger}a\rangle$ of the cavity field (dashed line) plotted for
(a) $N=3$ and (b) $N=5$.}\label{fig1}
\end{figure}

It is well known that $N$ qubits can be defined by the states of
$N$ spatially separated microcrystallites. The two most
interesting states for both experimentalists and theoreticians are
the no-exciton and one-exciton states denoted by $|0\rangle$ and
$|1\rangle$, respectively. So, we choose the computational basis
states of the qubit as $\{|0\rangle, |1\rangle\}$ for each
microcrystallite. If the cavity field is initially in the
single-photon state $|\psi(0)\rangle_{C} =a^{\dagger}
|0\rangle_{C}$, which now can successfully be prepared by the
experiment, and no exciton is initially in any microcrystallite,
then $|\Psi(0)\rangle=a^{\dagger}|0\rangle_{C}|0\rangle^{\otimes
N}$. Based on this initial condition, we interpolate the unit
operator $U^{\dagger}(t)U(t)$ into Eq. (\ref{eq:5}) and consider
the properties of the time-evolution operator
$U^{\dagger}(t)OU(t)=O(t)$ and $U(t)|0\rangle=|0\rangle$, the wave
function of the whole system can be obtained as follows
\begin{eqnarray}\label{eq:6}
|\Psi (t)\rangle &=& U(t)a^{\dagger }|0\rangle _{C}|0\rangle
^{\otimes
N}=a^{\dagger }(-t)|0\rangle _{C}|0\rangle ^{\otimes N}  \nonumber \\
&=&\Big[ a^{\dagger }(0)\cos
(G^{\prime}t)-i\sum_{j}f_{j}b_{j}^{\dagger }(0)\Big]
e^{-i\omega t}|0\rangle _{C}|0\rangle ^{\otimes N}  \nonumber \\
&=&-ie^{-i\omega t}|0\rangle _{C}\sum_{j}f_{j}|1\rangle
_{j}|0\rangle ^{\otimes (N-1)}\nonumber\\
&&+e^{-i\omega t}\cos (G^{\prime}t)|1\rangle _{C}|0\rangle
^{\otimes N},
\end{eqnarray}
which has been returned into the original frame, and
$|1\rangle_{j}|0\rangle^{\otimes (N-1)}$ means that $N-1$
microcrystallites have no excitons, and only one exciton is
excited by the cavity field in the $j$th microcrystallite.  We are
interested in the entanglement between two subsystems of the
excitons, such as, the $n$th and $m$th microcrystallites, then
after tracing out the cavity field and the degrees of freedom of
other $N-2$ microcrystallites, the reduced density operator for
this pair of qubits can be obtained as
\begin{eqnarray}\label{eq:7}
\rho (t) &=&f_{m}^{2}|10\rangle \langle 10|+f_{m}f_{n}|10\rangle
\langle 01|
\nonumber \\
&&+f_{n}^{2}|01\rangle \langle 01|+f_{m}f_{n}|01\rangle \langle
10|
\nonumber \\
&&+\Big\{ \cos ^{2}(G^{\prime}t)+\sum_{l\neq \{n,m\}}f_{l}^2\Big\}
|00\rangle \langle 00|.
\end{eqnarray}
The entanglement between two qubits can  mathematically be
described by using the concurrence~\cite{wt}. We assume a pair of
qubits whose density matrix  is $\rho_{12}$. Then the concurrence
of the density matrix  $\rho_{12}$ is defined as
\begin{equation}\label{eq:8}
 C=\max \{ \lambda_{1}-\lambda_{2}-\lambda_{3}-\lambda_{4},0
\}, \label{eq:c}
\end{equation}
where $\lambda_{1}$, $\lambda_{2}$, $\lambda_{3}$, and
$\lambda_{4}$, given in decreasing order, are the square roots of
eigenvalues for the matrix
\begin{equation}\label{eq:10}
M_{12}=\rho_{12}(\sigma_{1y}\otimes \sigma_{2y})
\rho^{*}_{12}(\sigma_{1y}\otimes \sigma_{2y}), \label{eq:m}
\end{equation}
with the Pauli matrix $$\sigma_{1y}=\sigma_{2y} =\left(
\begin{array}{cc}  0 &-i \\ i & 0\end{array}\right),$$ where the
asterisk denotes complex conjugation in the standard basis
$\{|00\rangle, |01\rangle, |10\rangle,  |11\rangle \}$, and
$\sigma_{1y}$  and $\sigma_{2y}$ are expressed in the same basis.
The entanglement of formation is a monotonically increasing
function of $C$; and $C=0$ ($C=1$) corresponds to an unentangled
state (maximally entangled state). The concurrence for the reduced
density operator (\ref{eq:7}) can be obtained using Eqs.
(\ref{eq:8}) and (\ref{eq:10}) as follows
\begin{equation}
C(t)=2f_{m}f_{n}=2g_{m}g_{n}\frac{\sin
^{2}(G^{\prime}t)}{G^{\prime 2}}.
\end{equation}
It is found that the concurrence $C$ periodically reaches its
maximum value, but the values of the concurrences are different
for different pairs, which means that the entanglements between
different pairs are different. The coupling constants between the
cavity field and microcrystallites determine the entanglement of
each pair. So, we can realize symmetric sharing of entanglement of
excitonic states in semiconductor microcrystallites only when all
microcrystallites have the same interaction with the cavity field,
e.g., $g_{1}=g_{2}=\cdots=g_{N}=g$, which may be obtained with the
development of the microfabrication technology in the near future.
Under this condition we can obtain the concurrence as
\begin{equation}\label{eq:ppp}
C(t)=\frac{2}{N}(1-\langle
a^{\dagger}a\rangle)=\frac{2}{N}\sin^2(Gt),
\end{equation}
with $G=g\sqrt{N}$, and the concurrence $C$ periodically reaches
its maximum value of $2/N$. Comparing the time evolution of the
concurrence and the average photon number $\langle
a^{\dagger}a\rangle$ of the cavity field, we can easily find that
when the average photon number is zero, the concurrence reaches
the maximal value of $2/N$ for any number $N$ of the
microcrystallites and vice versa. As an example for $N=3$ and $5$,
Fig.~\ref{fig1} clearly shows this point.
\noindent  Under the condition that all microcrystallites have the
same coupling  with the cavity field, when the concurrence reaches
its maximum values, the state of the microcrystallites system is
in the generalized $W$ state defined~\cite{xd} to be
$|W_{N}\rangle=N^{-1/2}(|10\cdots 0\rangle+|010\cdots
0\rangle+\cdots+|0\cdots 01\rangle)$ and the cavity field is in
the vacuum state, but in the anisotropic case, we cannot obtain
the  generalized $W$ state for any condition. We know that a
single exciton can be taken as a boson even for the quantum dots,
so the assumption of the microcrystallites is not necessary in the
case of the cavity field initially in the single-photon state.
 In the following two sections, we
will mainly focus on the isotropic interaction of the cavity field
and microcrystallites to discuss the entangled coherent excitonic states.

\section{PREPARATION OF ENTANGLED COHERENT EXCITONIC STATE}

There are other two interesting orthogonal states called the even
and odd coherent states (CS). These can be used as a robust qubit
encoding for a single bosonic mode subject to amplitude damping,
because the error caused by amplitude damping for this encoding
can easily be corrected by a standard three-qubit error-correction
circuit~\cite{pmj}. So, in this section, we will discuss how to
realize symmetric sharing of entanglement between any pair of
qubits defined by the even (odd) coherent excitonic states in
semiconductor microcrystallites. It is well known that one can
define the even CS as the zero-qubit state $|0\rangle$ and the odd
CS as the one-qubit state $|1\rangle$ to encode a CNOT quantum
gate~\cite{mw}, that is
\begin{mathletters}
\begin{eqnarray}
|0\rangle&=&N_{+}(|\alpha\rangle+|-\alpha\rangle)\label{eq:12a}, \\
|1\rangle&=&N_{-}(|\alpha\rangle-|-\alpha\rangle)\label{eq:12b},
\end{eqnarray}
\end{mathletters}
with the normalization constants $N_{\pm}\!=\! (2\pm
2e^{-2|\alpha|^2}) ^{-1/2}$ and $|\pm\alpha\rangle=
\exp(-|\alpha|^2/2) \sum_{n=0}^{\infty}
[(\pm\alpha)^n/\sqrt{n!}]|n\rangle$ are coherent states of a
bosonic annihilation operator, e.g., the coherent states of the
annihilation operator $a$ for the cavity field. The even and odd
coherent superpositions of the photon states in cavity quantum
electrodynamics and those of motional states of trapped ions can
be created by experimentalists~\cite{haroche} over the past
several years. So, we can assume that the cavity field is
initially either in the odd CS or in the even CS, and there are no
excitons in any microcrystallite. In order to realize the
symmetric sharing of entanglement, we assume that all
microcrystallites have the same interaction with the cavity field,
then the wave function of the whole system can be written by the
factorization of the wave function~\cite{sun} as follows
\begin{eqnarray}\label{eq:11}
|\Psi(t)\rangle &=& N_{\pm}U(t)[|\alpha\rangle_{C} \pm
|-\alpha\rangle_{C}] |0\rangle^{\otimes N}\nonumber\\
 &=& N_{\pm}(|\alpha
u(t)\rangle_{C} |v(t)\alpha \rangle^{\otimes N} \nonumber\\
 &&\pm |-\alpha u(t)\rangle_{C}|-v(t)\alpha \rangle^{\otimes N}),
\end{eqnarray}
with $u(t)\!=\!\cos(G t)e^{-i\omega t}$ and ${v(t)\!=\!-i[\sin
(Gt)/\sqrt{N}]e^{-i\omega t}}$ and the same coupling constants
between microcrystallites and the cavity field are taken. We find
that all excitonic coherent states $|v(t)\alpha\rangle$ in
microcrystallites evolve periodically with time evolution and
their maximal amplitudes are $1/\sqrt{N}$ times the amplitude
$|\alpha|$ of the coherent cavity field. We are interested in the
pairwise entanglement in the system of $N$ microcrystallites.
After tracing out the cavity field and other degrees of freedom
for $N-2$ microcrystallites, the reduced density operator for any
pair can be expressed as
\begin{eqnarray}\label{eq:13}
\rho(t)&=&N^{2}_{\pm}\{(|v(t)\alpha\rangle\langle
v(t)\alpha|)^{\otimes 2}+(|-v(t)\alpha\rangle\langle
-v(t)\alpha|)^{\otimes 2} \nonumber\\
&& \pm  P(t)(|v(t)\alpha\rangle\langle -v(t)\alpha|)^{\otimes
2}\nonumber\\
&&\pm  P(t)(|-v(t)\alpha\rangle\langle v(t)\alpha|)^{\otimes 2}\},
\end{eqnarray}
where $P(t)=\langle -u(t)\alpha|u(t)\alpha\rangle(\langle
-v(t)\alpha|v(t)\alpha\rangle)^{N-2}= \exp[-2|\alpha|^2+4
|\alpha|^2\sin^2(Gt)/N]$. We choose the time-dependent even and
odd CS as the basis $\{|\widetilde{0}
\rangle,|\widetilde{1}\rangle\}$ for each qubit in every
microcrystallite  as follows~\cite{kim}
\begin{mathletters}
\begin{eqnarray}
&&|\widetilde{0}\rangle=N_{+}(t)(|v(t)\alpha\rangle+|-v(t)\alpha\rangle),\\
&&|\widetilde{1}\rangle=N_{-}(t)(|v(t)\alpha\rangle-|-v(t)\alpha\rangle),
\end{eqnarray}
\end{mathletters}
where $N_{\pm}(t)$ are the normalization constants defined as
$N_{\pm}(t)=(2\pm 2e^{-2|\alpha|^{2}\sin^2(Gt)/N})^{-1/2}$. Then
the reduced density operator $\rho(t)$ can be given, in the basis
$\{|\widetilde{0}\hspace{0.5mm}\widetilde{0}\rangle,
|\widetilde{0}\hspace{0.5mm} \widetilde{1}\rangle,
|\widetilde{1}\hspace{0.5mm} \widetilde{0}\rangle,|\widetilde{1}
\hspace{0.5mm}\widetilde{1}\rangle\}$, in the following form
\begin{eqnarray}\label{eq:16}
\rho(t)&=&\frac{N^2_{\pm}(1 \pm
P(t))}{8N^{4}_{+}(t)}|\widetilde{0}\hspace{0.5mm}
\widetilde{0}\rangle\langle \widetilde{0}\hspace{0.5mm}
\widetilde{0}|+\frac{N^{2}_{\pm}(1 \pm P(t))}{8N^{4}_{-}(t)}
|\widetilde{1}\hspace{0.5mm}\widetilde{1}\rangle\langle
\widetilde{1}\hspace{0.5mm}\widetilde{1}|
 \nonumber\\
&&+\frac{N^{2}_{\pm}(1 \mp P(t))}{8N^{2}_{+}(t)N^{2}_{-}(t)} \left
\{ |\widetilde{0}\hspace{0.5mm}\widetilde{1}\rangle\langle
\widetilde{0}\hspace{0.5mm}\widetilde{1}|+
|\widetilde{0}\hspace{0.5mm}\widetilde{1}\rangle\langle
\widetilde{1}\hspace{0.5mm}\widetilde{0}|
+|\widetilde{1}\hspace{0.5mm}\widetilde{0}\rangle\langle
\widetilde{0}\hspace{0.5mm}\widetilde{1}|\right.\nonumber\\
&&\left.+ |\widetilde{1}\hspace{0.5mm}\widetilde{0}\rangle\langle
\widetilde{1}\hspace{0.5mm}\widetilde{0}| \right \}
+\frac{N^{2}_{\pm}(1 \pm P(t))}{8N^{2}_{+}(t)N^{2}_{-}(t)}
 \nonumber\\
&&\times\left\{|\widetilde{0} \hspace{0.5mm}
\widetilde{0}\rangle\langle
\widetilde{1}\hspace{0.5mm}\widetilde{1}|+
|\widetilde{1}\hspace{0.5mm}\widetilde{1}\rangle\langle
\widetilde{0}\hspace{0.5mm}\widetilde{0}| \right \}.
\end{eqnarray}
Following the same steps as for Eqs. (\ref{eq:8}) and
(\ref{eq:10}), we obtain the concurrence corresponding to Eq.
(\ref{eq:16}) as
\begin{eqnarray}\label{15}
&&C_{\pm}(t)=\frac{e^{4|\alpha|^2\sin^2(Gt)/N}-1}{e^{2|\alpha|^2}
\pm 1},
\end{eqnarray}
where $+(-)$ means that the cavity field is initially in the even
(odd) CS. It is also found that the concurrence periodically
evolves. Although there is no  simple analytical expression
between the concurrence and the average photon number $\langle
a^{\dagger}a\rangle$ of the cavity field as Eq. (\ref{eq:ppp}), we
find that if the average photon number of the cavity field
\begin{equation}
\langle a^{\dagger}a\rangle_{\pm} =|\alpha|^2\cos^2(Gt)\frac{1\mp
e^{-2|\alpha|^2}}{1 \pm e^{-2|\alpha|^2}}
\end{equation}
is zero, then the value of the concurrence reaches maximum and
vice versa. So, for the fixed number $N$ of the microcrystallites
and the intensity $|\alpha|^2$ of the cavity field, the
relationship of the time evolution between the concurrence and the
average number of the cavity field is analogous to
Fig.~\ref{fig1}.

\begin{figure}
\includegraphics[width=8cm]{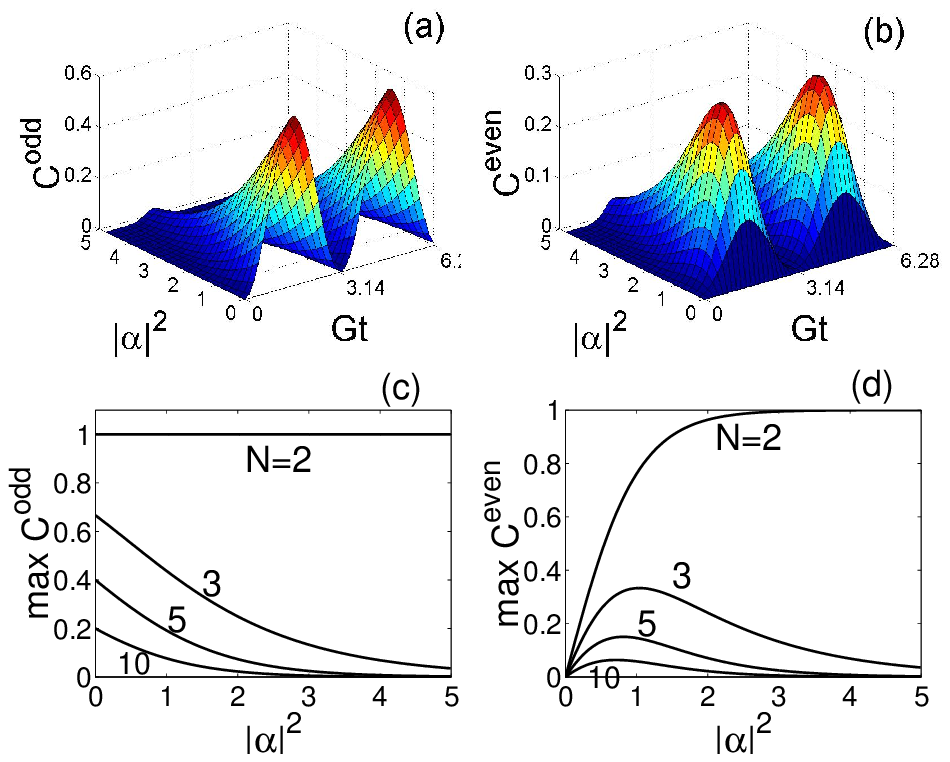}
\caption[]{The concurrences are plotted as a function of time  and
intensity $|\alpha|^2$ of the cavity field when $N=3$ for the
cavity field initially in (a) the odd CS and (b) the even CS. The
maximal concurrences  vs $|\alpha |^{2}$ for $N$=2, 3, 5, and 10
at optimum evolution times  are plotted for the cavity field
initially in (c) the odd CS and (d) the even CS.}\label{fig2}
\end{figure}

It is very clear that the values of the concurrence (\ref{15})
depend on both the number $N$ of the microcrystallites and the
intensity $|\alpha|^2$ of the cavity field. The maximal
concurrence with the cavity field initially in even CS or odd CS
decreases with the increase of the microcrystallite's number $N$
when the intensity $|\alpha|^2$ of the cavity field is fixed. It
is because that $\exp\{4|\alpha|^2\sin^2(Gt)/N\}$ in Eq.
(\ref{15}) is a decreasing function of the number $N$, so a larger
number $N$ corresponds to a smaller concurrence. In the following,
we have plotted Fig.~\ref{fig2} to show  the time evolution of the
concurrences, and the relationship between the concurrence and the
intensity $|\alpha|^2$ for different microcrystallite number $N$.
We find that the concurrence $C_{-}(t)\equiv C^{\rm odd}(t)$
periodically reaches its maximal value at the evolution times $
Gt=(2n+1)\pi/2$ \thinspace ($n=0,1,...$), and these points
approach the upper bound value of $2/N$,  which has been
illustrated in Fig.~\ref{fig2}(a) and Fig.~\ref{fig2}(c) by an
example for $N=3$, when $|\alpha|^{2}\rightarrow 0$. It is because
\linebreak
\noindent the odd CS of the cavity field is reduced to the
single-photon state when $|\alpha|^2\rightarrow 0$, so the
concurrence $C_{-}(t)$ with initially the odd CS is reduced to Eq.
(\ref{eq:ppp}) and approaches upper bound value of $2/N$.
Figure~\ref{fig2}(c) also shows the variation for the maximal
value of the concurrence with the intensity $|\alpha|^2$ of the
cavity field with different number $N$ of the microcrystallites,
we find that the maximal values of the concurrence decrease with
the increase of the intensity $|\alpha|^2$, except that $N=2$.
However, when $|\alpha|^{2}\rightarrow 0$, the even CS of the
cavity field is reduced to the vacuum state, the concurrence
$C_{+}(t)\equiv C^{\rm even}(t)$ with initially the even CS tends
to zero. The maximal values of $C_{+}(t)$ increase with the
increase of the intensity $|\alpha|^{2}$, but when the intensity
$|\alpha|^{2}$ is greater than a threshold value, which is
determined by
\begin{equation}
N\!=\!\frac{4|\alpha |^{2}\cosh |\alpha |^{2}}{|\alpha
|^{2}e^{|\alpha |^{2}}+\cosh |\alpha |^{2}W\{-|\alpha |^{2}{\rm
sech}|\alpha |^{2}e^{-|\alpha |^{2}\tanh |\alpha |^{2}}\}}
\label{N18}
\end{equation}
for given number $N$ ($N>2$), where $W\{z\}$ is the product log
function defined as the solution for $w$ of $z=we^{w}$, then the
concurrence gradually tends to zero. The maxima of $C_{+}(t)$ are
reached at $Gt=(2n+1)\pi/2$ if  $N$ and $|\alpha |^{2}$ satisfy
Eq.~(\ref{N18}). We  can also find when $|\alpha|^2$ is large
enough, such that $|\alpha|^2 \approx 4$ when $N\geq 4$ and
$|\alpha|^2 \approx 6$ when $N=3$ [see Figs. \ref{fig2}(a) and
\ref{fig2}(b)], where bosonic approximation for excitons is still
good, then  $|\alpha v(t)\rangle$ is approximately  orthogonal to
$|-\alpha v(t)\rangle$  when $u(t)=0$, that is, $\langle
-v(t)\alpha|v(t)\alpha\rangle \approx 0$. Under such condition, we
can redefine two approximately orthogonal states
$|1\rangle=|\alpha v(t)\rangle$ and $|0\rangle=|-\alpha
v(t)\rangle$ as one qubit state and zero qubit state, then
Eq.~(\ref{eq:16}) for the reduced density operator of two
subsystems of $N$ ($N>2$) microcrystallites  can be simplified as
\begin{figure}
\includegraphics[width=8cm]{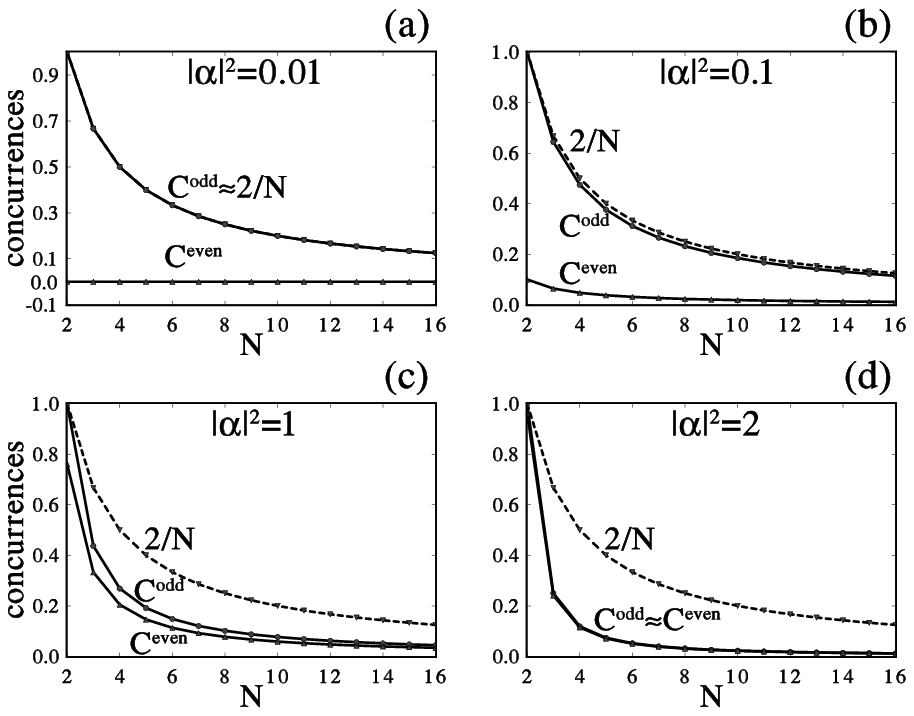}
\caption[]{The maximum values of the concurrences are depicted as
a function of the number $N$ of semiconductor microcrystallites
for the different intensities of the cavity field with initially
the odd or even CS for (a) $|\alpha|^2=0.01$, (b)
$|\alpha|^2=0.1$, (c) $|\alpha|^2=1$, and (d) $|\alpha|^2=2$, and
the dashed curve corresponds to the maximum possible concurrence,
given by $2/N$, between any pair of qubits.}\label{fig3}
\end{figure}
\begin{equation}
\rho\approx\frac{1}{2}[|00\rangle\langle00|\pm|11\rangle\langle 11|],
\end{equation}
which means that no entanglement appears for each pair of
microcrystallites, then the concurrences are zero for cavity field
initially in odd CS or even CS. So when the cavity field is
initially in the even CS, the points of the maximal values for the
concurrence must be between $0$ and $|\alpha|^2$, with condition
$\langle -v(t)\alpha|v(t)\alpha\rangle \approx 0$. These points
are determined by Eq.~(\ref{N18}).

The optimum values of $|\alpha |^{2}$ and $N$ maximizing
$C_{+}(t)$  obtained from Eq. (\ref{N18}) and checked directly by
numerical maximization of Eq. (\ref{15}) are, e.g., as follows:
$|\alpha|^2 = 3/2 \ln(2)\approx 1.04$ for $N=3$, and $|\alpha|^2 =
\ln(1 + \sqrt{2})\approx 0.88$ for $N=4$. While for $N=5$ the
maximum is at $|\alpha|^2 = \ln[16/9 + 10^{1/3} 20/27 + 10^{2/3}
10/27]/2 \approx 0.81$.  For $N$=6 and 8, analytical expressions
can also be found. However, for other cases (i.e., $N=7$ and $N\ge
9$) there are no compact-form analytical formulas for $|\alpha|^2$
corresponding to the maximal concurrence. The above numerical
calculations can be seen from Figs. \ref{fig2}(b) and
\ref{fig2}(d). Figure~\ref{fig2} also shows when the number $N$ of
the microcrystallites is equal to 2, the maximally entangled
states can be prepared  for any intensity of the cavity field with
initially the odd CS, but for the cavity field with initially the
even CS, we can approximately prepare a maximally entangled state
when the average photon number $|\alpha|^2$ is slightly larger
than 1, e.g.,  $|\alpha|^2 \geq 3$ [see Fig. \ref{fig2} (d)].

As a comparison, we numerically show that the maximal values of
the concurrence in the case of the cavity field initially in the
odd or even CS for the different number $N$ of the
microcrystallites and the different intensity $|\alpha|^2$ of the
cavity field from Fig.~\ref{fig3}(a) to Fig.~\ref{fig3}(d).
Figure~\ref{fig3} shows that the maximal values of the
concurrences with initially the odd or even CS of the cavity field
approach each other with increasing the intensity of the cavity
field, but they are lower than the upper bound values of $2/N$ for
the concurrence. From Fig.~\ref{fig3}(a), we can also find that
the concurrence with the cavity field initially in the odd or even
CS gradually approaches its maximal possible values of $2/N$ or
zero when the intensity of the cavity field tends to zero, that
is, $|\alpha|^2\rightarrow 0$.

\section{DECAY OF THE ENTANGLED EXCITON STATES}

The quantum computation and the quantum information take their
power from the superpositions and entanglement of the quantum
states, however, the necessary coupling of the system to the
environment tends to destroy this coherence and reduces the degree
of the entanglement with the time evolution of the total system.
So, in this section, we will discuss the decay of the entangled
exciton states. For simplification of our discussion, we will
model the interaction between the system and environment as
follows. We assume that there is no interaction between the cavity
field and environment. The dissipation of the system energy comes
from the interaction of the excitons in the microcrystallites with
environment, here modeled as thermal radiation fields at zero
temperature. We limit our discussion to the completely identical
$N$ microcrystallites, and the coupling constants between
microcrystallites and cavity field are the same. Under the above
assumptions, the Hamiltonian for the system, environments, and
their interactions can be written as
\begin{eqnarray}
&&H=\hbar\omega a^{\dagger}a+\hbar\omega\sum_{j=1}^{N} b^{\dagger}_{j}
b_{j}+\hbar g\sum_{j=1}^{N}
(a^{\dagger}b_{j}+ab^{\dagger}_{j})\nonumber\\
&&+
\hbar\sum_{j=1}^{N}\sum_{k}\omega_{j,k}a^{\dagger}_{j,k}a_{j,k}
+\hbar\sum_{j=1}^{N}\sum_{k}g_{j,k}(a^{\dagger}_{j,k}b_{j}+a_{j,k}
b^{\dagger}_{j}),
 \nonumber\\
 \end{eqnarray}
where $a_{j,k}$($a^{\dagger}_{j,k}$) are annihilation (creation)
operators of radiation fields with frequency $\omega_{j,k}$., and
$g_{j,k}$ are coupling constant between the $j$th microcrystallite
and radiation fields. For simplicity, we assume that all $g_{j,k}$
are independent of the microcrystallite size. We assume that each
microcrystallite separately interacts with the environment, but
the dissipative dynamics is same for all the microcrystallites.
The latter assumption is not necessary but only simplifies the
degree of algebra complexity for calculation.

We can obtain the Heisenberg equations of motion for each operator
as follows
\begin{mathletters}
\begin{eqnarray}
\frac{\partial B_{j}}{\partial t}&=& -i g A-
i\sum_{k}g_{j,k}A_{j,k}e^{-i(\omega_{j,k}-\omega)t},\label{18a}\\
\frac{\partial A}{\partial t}&=&-i
g\sum_{j=1}^{N}B_{j},\label{18b}\\
\frac{\partial A_{j,k}}{\partial t}&=& -i
g_{j,k}B_{j}e^{i(\omega-\omega_{j,k})t} , \label{eq:18c}
\end{eqnarray}
\end{mathletters}
where the transformations $a(t)=A(t)e^{-i\omega t}$,
$b_{j}(t)=B_{j}(t)e^{-i\omega t}$, and
$a_{j,k}(t)=A_{j,k}(t)e^{-i\omega_{j,k} t}$ are made. From
Eq.~(\ref{eq:18c}), we have
\begin{equation}
A_{j,k}(t)=A_{j,k}(0)-ig_{j,k}\int_{0}^{t}{\rm d} t^{\prime}
B_{j}(t^{\prime})e^{i(\omega-\omega_{j,k})t^{\prime}}.\label{eq:19}
\end{equation}
We replace $A_{j,k}(t)$ in Eqs. (\ref{18a}) by (\ref{eq:19}), then
obtain the new equation as
\begin{eqnarray}
\frac{\partial B_{j}}{\partial t}&=& -i g A-i
\sum_{k}g_{j,k}A_{j,k}(0) e^{-i(\omega_{j,k}-\omega)t}\nonumber\\
&&-\sum_{k}|g_{j,k}|^{2} \int_{0}^{t}{\rm d} t^{\prime}
B_{j}(t^{\prime})e^{-i(\omega_{j,k}-\omega)(t-t^{\prime})}.
\label{eq:20}
\end{eqnarray}
We can apply the Laplace transform and the Wigner-Weisskopf
approximation~\cite{lou} to Eqs.~(\ref{18b}) and (\ref{eq:20}), so
that we have the solution of the cavity field as
\begin{equation}
A(t)=u^{\prime}(t)a(0)-i\sum_{j}v^{\prime}(t)b_{j}(0)
+\sum_{j}\sum_{k}v_{j,k}(t)a_{j,k}(0),
\end{equation}
where conditions $A(0)=a(0)$, $B_{j}(0)=b_{j}(0)$, and
$A_{j,k}(0)=a_{j,k}(0)$ were used, and
\begin{mathletters}
\begin{eqnarray}
u^{\prime }(t) &=&e^{-(\gamma/4) t}\left( \frac{\gamma }{4\delta
}\sin (\delta
t)+\cos (\delta t)\right),  \\
v^{\prime }(t) &=& \frac{g}{\delta }e^{-(\gamma/4) t}
\sin (\delta t), \label{22b} \\
\delta &=&\sqrt{Ng^{2}-(\gamma/4) ^{2}},
\end{eqnarray}
\end{mathletters}
where the small Lamb frequency shift is neglected and the decay
rate $\gamma=2\pi \rho(\omega_{0})|g(\omega_{0})|^2$.  We also use
the former assumptions under which all microcrystallites are the
same and have the same dissipative dynamics so that the decay
rates $\gamma$ of each microcrystallite and the functions
$v^{\prime}(t)$ of each term including operators $b_{j}(0)$ are
the same. When the environment is considered,  the qubits for each
microcrystallite should be redefined as
\begin{mathletters}
\begin{eqnarray}
&&|\widetilde{0}\rangle=M_{+}(t)
[|v^{\prime}(t)\alpha\rangle+|-v^{\prime}(t)\alpha\rangle],\\
&&|\widetilde{1}\rangle=M_{-}(t)
[|v^{\prime}(t)\alpha\rangle-|-v^{\prime}(t)\alpha\rangle],
\end{eqnarray}
\end{mathletters}
with the normalization ${M_{\pm}(t)=[2\pm
2e^{-2|v^{\prime}(t)\alpha|^2}]^{-1/2}}$.  Now, we will
investigate the decay when the cavity field is initially in the
even and odd CS, but no excitons are initially in any
microcrystallite. After tracing out the degrees of the
environments and other $N-2$ microcrystallites for the
time-dependent wave function of the whole system, which can also
be obtained using the factorized form of the wave function, we get
the reduced density operator for any two qubits as
\begin{figure}
\includegraphics[width=8cm]{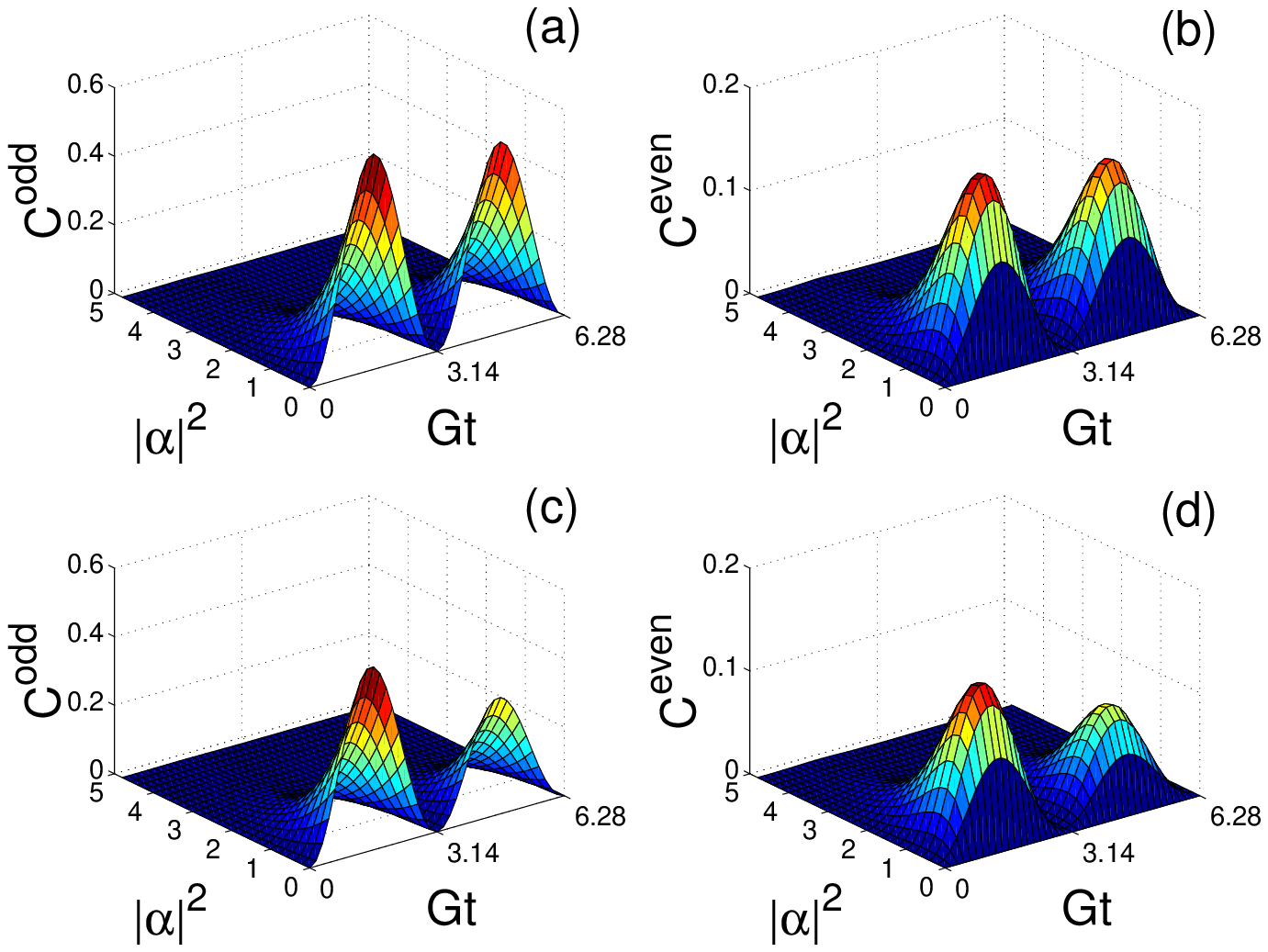}
\caption[]{The time evolution of the concurrences depicted with
the number $N=3$ of semiconductor microcrystallites for
${\gamma/g}=0.13$ (a, b) and  ${\gamma/g}=0.5$ (c, d) when the
cavity field is initially in the odd  CS (a, c) or in the even CS
(b, d).}\label{fig4}
\end{figure}
\begin{eqnarray}
\rho^{\prime}_{\pm}(t)&=&\frac{N^2_{\pm}(1 \pm
P^{\prime}(t))}{8M^{4}_{+}(t)}|\widetilde{0}\hspace{0.5mm}
\widetilde{0}\rangle\langle
\widetilde{0}\hspace{0.5mm}\widetilde{0}|
 \nonumber\\
&&+\frac{N^{2}_{\pm}(1 \pm P^{\prime}(t))}{8M^{4}_{-}(t)}
|\widetilde{1}\hspace{0.5mm}\widetilde{1}\rangle\langle
\widetilde{1}\hspace{0.5mm}\widetilde{1}|
 +\frac{N^{2}_{\pm}(1 \mp
P^{\prime}(t))}{8M^{2}_{+}(t)M^{2}_{-}(t)}
 \nonumber\\ &&\times
\{ |\widetilde{0}\hspace{0.5mm}\widetilde{1}\rangle\langle
\widetilde{0}\hspace{0.5mm}\widetilde{1}|+
|\widetilde{0}\hspace{0.5mm}\widetilde{1}\rangle\langle
\widetilde{1}\hspace{0.5mm}\widetilde{0}|
+|\widetilde{1}\hspace{0.5mm}\widetilde{0}\rangle\langle
\widetilde{0}\hspace{0.5mm}\widetilde{1}|+
|\widetilde{1}\hspace{0.5mm}\widetilde{0}\rangle\langle
\widetilde{1}\hspace{0.5mm}\widetilde{0}| \}
 \nonumber\\
&&+\frac{N^{2}_{\pm}(1 \pm
P^{\prime}(t))}{8M^{2}_{+}(t)M^{2}_{-}(t)}
\{|\widetilde{0}\hspace{0.5mm}\widetilde{0}\rangle\langle
\widetilde{1}\hspace{0.5mm}\widetilde{1}|+
|\widetilde{1}\hspace{0.5mm}\widetilde{1}\rangle\langle
\widetilde{0}\hspace{0.5mm}\widetilde{0}|  \},\label{24}
\end{eqnarray}
with $P^{\prime}(t)=\exp\left[-2|\alpha|^2\left(1-2|
v^{\prime}(t)|^2\right)\right]$. Then the concurrences
$C^{\prime}_{\pm}(t)$ corresponding to Eq.~(\ref{24}) can be
obtained as
\begin{equation}\label{26}
C^{\prime}_{\pm}(t)=\frac{e^{4|\alpha v^{\prime }(t)|^{2}}-1}{
e^{2|\alpha |^{2}}\pm1},
\end{equation}
where $v^{\prime}(t)$ is determined by Eq.~(\ref{22b}). We find
that the concurrence (\ref{15}) is modified and becomes of the
form (\ref{26}) after the effect of the environment is taken into
account. Now we will make further approximation. We assume that
the couplings of the cavity field with microcrystallites are
stronger than  the decay of the exciton, i.e., $g\gg\gamma$. In
this case we can approximately obtain the concurrence
$C^{\prime}_{\pm}(t)$ as
\begin{equation}\label{28}
C^{\prime}_{\pm}(t)\approx \frac{\exp \left[ (4|\alpha
|^{2}/N)\sin ^{2}(gt\sqrt{N})e^{-(\gamma
/2)t}\right]-1}{e^{2|\alpha |^{2}}\pm 1}.
\end{equation}
Equation~(\ref{28}) shows that the entanglement between any pair
of qubits decays in an oscillating form. It also shows that the
increase of the  microcrystallite number $N$ results in the
decrease of the concurrence when the decay rate $\gamma$ and the
intensity $|\alpha|^2$ are fixed. For convenience of discussion,
we can rescale time in Eq.~(\ref{28}) as $t^{\prime}=Gt$, then we
can find that if the coherent intensity $|\alpha|^2$ of the cavity
field and the number $N$ of microcrystallites are given, then the
larger ratio between decay rate $\gamma$ and the coupling constant
$g$ corresponds to the faster reduction of the concurrence of the
entangled qubits.  But if the ratio of $\gamma$ and $g$ is given
and the number $N$ is fixed, then the higher intensity of the
cavity field corresponds to the smaller concurrence. As an
example, Fig.~\ref{fig4} plots the variation of concurrence with
$N=3$ for a reasonably good cavity $\gamma/g=0.13$ [in
Fig.~\ref{fig4} (a) and (b)]~\cite{Ima}, or for a bad cavity
$\gamma/g=0.5$ [in Fig.~\ref{fig4} (c) and (d)] according to
Eq.~(\ref{26}). Figure~\ref{fig4} clearly demonstrates our above
discussions.

\section{CONCLUSIONS}

We have studied an excitonic-state implementation of the
multiparticle entanglement  based on $N$ spatially separated
semiconductor microcrystallites. The interaction among the
microcrystallites is mediated by  a single-mode cavity field. We
find that the entanglement (measured by the concurrence) between
any pair of qubits that are defined by the excitonic number states
(vacuum and a single-exciton states) or the coherent excitonic
states (odd and even CS), depends on the interaction between the
cavity field and the semiconductor microcrystallites. The
entanglement between any pairs is different from one  another for
the anisotropic case. When all microcrystallites  have the same
interaction with the cavity field, the maximal degree of the
entanglement between any pair of qubits is the same. This
condition can, probably, be satisfied with the development of the
fabrication techniques for quantum dots and the semiconductor
microcavity quantum electrodynamics. So, the symmetric sharing of
the entanglement between any pair of $N$ qubits in such a system
is realizable only when the interaction between $N$ spatially
separated semiconductor microcrystallites and the cavity field is
isotropic. Under the isotropic-interaction condition, when the
excitonic system reaches maximal entanglement, all photons in the
cavity are transformed into the excitons in the system of the
semiconductor microcrystallites. The generalized $W$ state and the
maximal degree $2/N$ of entanglement can be obtained for the
cavity field initially in the single-photon state. But if the
cavity field is initially in the odd or even CS, we cannot obtain
the maximal degree of entanglement $2/N$ only except the special
case where the cavity field is initially in the odd CS and there
are two microcrystallites in the cavity~\cite{Liu}.

We have also investigated the decay of any pair of the entangled
qubits defined by the odd and even CS. The entanglement between
any pair of qubits decreases because of the dissipation of the
system energy to the environment. If the coherent intensity of the
cavity field and the number of  microcrystallites are given, then
with the rescaled time,  the larger ratio between the decay rate
$\gamma$ and the coupling constant $g$ corresponds to the faster
reduction of the concurrence of the entangled qubits. For the
given ratio between the decay rate $\gamma$ and the coupling
constant $g$ and the coherent intensity of the cavity field, the
increase of the  microcrystallite number $N$ results in the
decrease of the concurrence. But if the ratio $\gamma/g$ is given,
and the number $N$ is fixed, then the higher intensity of the
cavity field corresponds to the smaller concurrence. Practically,
the quality of the entanglement can be improved with the
appearance of the new processing techniques and the ultrahigh
finesse cavities~\cite{Ima}. Finally, we should point out that our
discussion is limited to the preparation of the entangled coherent
excitonic states, but we cannot control them using this model.

\section*{ACKNOWLEDGMENTS}

The authors are grateful to Yoshiro Hirayama and \c{S}ahin K.
\"Ozdemir for most helpful discussions. One of authors (Yu-xi Liu)
is supported by the Japan Society for the Promotion of Science
(JSPS).


\vspace{1mm} {\setlength{\fboxsep}{1pt}
\begin{center}
\framebox{\parbox{0.75\columnwidth}{%
\begin{center}
published in\\ {\em Physical Review A} {\bf 66}, 062309 (2002)\\
and selected to\\ {\em  Virtual J. Quantum Information}\\  {\bf 2}
(2002) Issue 12\\ {\tt (http://www.vjquantuminfo.org)}
\end{center}}}
\end{center}

\end{document}